# APPLESOSS: A Producer of ProfiLEs for SOSS. Application to the NIRISS SOSS Mode

Michael Radica[1] , Loïc Albert[1] , Jake Taylor[1] , David Lafrenière[1] , Louis-Philippe Coulombe[1] ,
Antoine Darveau-Bernier[1] , René Doyon[1] , Neil Cook[1] , Nicolas Cowan[2,3] , Néstor Espinoza[4,5] , Doug Johnstone[6,7] ,
Lisa Kaltenegger[8,9] , Caroline Piaulet[1] , Arpita Roy[4,5] , and Geert Jan Talens[1,10]
[1] Institut de Recherche sur les Exoplanètes and Département de Physique, Université de Montréal, 1375 Avenue Thérèse-Lavoie-Roux, Montréal, QC, H2V 0B3,
Canada; michael.radica@umontreal.ca
[2] Department of Earth & Planetary Sciences, McGill University, 3450 rue University, Montréal, QC H3A 0E8, Canada
[3] Department of Physics, McGill University, 3600 rue University, Montréal, QC H3A 2T8, Canada
[4] Space Telescope Science Institute, 3700 San Martin Drive, Baltimore, MD 21218, USA
[5] Department of Physics and Astronomy, Johns Hopkins University, 3400 N Charles St, Baltimore, MD 21218, USA
[6] NRC Herzberg Astronomy and Astrophysics, 5071 West Saanich Rd, Victoria, BC, V9E 2E7, Canada
[7] Department of Physics and Astronomy, University of Victoria, Victoria, BC, V8P 5C2, Canada
[8] Carl Sagan Institute, Cornell University, Ithaca, NY 14853, USA
[9] Department of Astronomy, Cornell University, Ithaca, NY 14853, USA
[10] Department of Astrophysical Sciences, Princeton University, 4 Ivy Lane, Princeton, NJ 08544, USA
Received 2022 July 7; accepted 2022 September 20; published 2022 October 11

## Abstract

The SOSS mode of the Near Infrared Imager and Slitless Spectrograph instrument is poised to be one of the workhorse modes for exoplanet atmosphere observations with the newly launched James Webb Space Telescope (JWST). One of the challenges of the SOSS mode, however, is the physical overlap of the first two diffraction orders of the G700XD grism on the detector. Recently, the ATOCA algorithm was developed and implemented as an option in the official JWST pipeline, as a method to extract SOSS spectra by decontaminating the detector—that is, separating the first and second orders. Here, we present A Producer of ProfiLEs for SOSS (APPLESOSS), which generates the spatial profiles for each diffraction order upon which ATOCA relies. We validate APPLESOSS using simulated SOSS time series observations of WASP-52 b, and compare it to ATOCA extractions using two other spatial profiles (a best and worst case scenario on-sky), as well as a simple box extraction performed without taking into account the order contamination. We demonstrate that APPLESOSS profiles retain a high degree of fidelity to the true underlying spatial profiles, and therefore yield accurate extracted spectra. We further confirm that the effects of the order contamination for relative measurements (e.g., exoplanet transmission or emission observations) is small—the transmission spectrum obtained from each of our four tests, including the contaminated box extraction, is consistent at the ∼1σ level with the atmosphere model input into our noiseless simulations. We further confirm via a retrieval analysis that the atmosphere parameters (metallicity and C/O) obtained from each transmission spectrum are consistent with the true underlying values.

*Unified Astronomy Thesaurus concepts:* Transmission spectroscopy (2133); Exoplanets (498); Exoplanet atmospheres (487); Astronomy data analysis (1858); Astronomical techniques (1684)

## 1. Introduction

In the past few decades, the study of exoplanets has evolved explosively from the detection of planets around distant stars (e.g., Mayor & Queloz 1995) to in-depth studies of the composition and dynamics of their atmospheres (e.g., Benneke et al. 2019; Welbanks et al. 2019; Boucher et al. 2021). Exoplanet atmosphere studies have generally relied upon space-based observations using the Hubble (HST) and Spitzer Space Telescopes; although the recent advent of ultra-stable ground-based spectrographs has enabled atmosphere studies to be performed from the ground as well (e.g., Brogi et al. 2016; Boucher et al. 2021; Line et al. 2021). Despite not being originally designed for exoplanet science, HST and Spitzer continue to be invaluable tools for exoplanet astronomers and have provided tantalizing glimpses into the compositional diversity of transiting exoplanets (Sing et al. 2016; Welbanks et al. 2019).

The recently launched James Webb Space Telescope (JWST) is the natural successor to HST and Spitzer. The increased wavelength coverage and spectral resolution of the JWST will facilitate a deeper understanding of exoplanet atmospheres than was possible with either of its predecessors (Greene et al. 2016), and each of the four instruments on board have observing modes specifically tailored to time series observations (TSOs) of exoplanets. Indeed, the Transiting Exoplanet Community Early Release Science program has already begun to demonstrate the unique insights that JWST will provide into the atmospheres of distant worlds (The JWST Transiting Exoplanet Community Early Release Science Team et al. 2022).

### 1.1. SOSS Order Contamination

The Single Object Slitless Spectroscopy (SOSS) mode (L. Albert et al. 2022, in preparation) of the Near Infrared Imager and Slitless Spectrograph (NIRISS) instrument (R. Doyon et al. 2022, in preparation) is poised to be one of the workhorse observing modes for exoplanet atmosphere studies. Its wide wavelength coverage (0.6–2.8 $\mu$m) extends the furthest to the blue of any instrument on board the JWST;[11] capturing not only the prominent near-infrared (NIR) water bands (Kreidberg et al. 2015; Sing et al. 2016), but also critical information on scattering slopes, which provide an anchor for the interpretation of the rest of a planet's NIR spectrum (Lecavelier des Etangs et al. 2008), as well as alkali metals such as potassium (Welbanks et al. 2019).

However, the extraction of spectral information from SOSS observations presents some challenge. The full wavelength range of the SOSS mode is split between two diffraction orders, with the first order covering the 0.9–2.8 $\mu$m range, and the second order covering 0.6–1.4 $\mu$m. After cross dispersion by the G700XD grism, the first and second orders partially overlap on the detector (e.g., Figures 1, 2). Red-wards of ∼2.4 $\mu$m, the first order is contaminated by the ∼1.1–1.4 $\mu$m regime of the second order. This physical overlap (or contamination; the two terms will be used interchangeably in this work, and are different from the wavelength overlap between the two orders; i.e., the 0.9–1.4 $\mu$m range common to the wavelength domain of both orders) complicates standard box extraction techniques. Darveau-Bernier et al. (2022) showed that the effects of this contamination depend on the stellar type of the target and can be significant for measurements of absolute flux. However, they are likely to be small (∼1% of the spectral feature amplitude) for relative measurements such as exoplanet atmosphere observations. In the interest of maximizing the scientific value of the JWST and the NIRISS instrument, it is important to properly treat this contamination during the extraction of SOSS spectra.

Recently, Darveau-Bernier et al. (2022, hereafter DB22) presented the ATOCA algorithm to deal with order contamination in the SOSS mode. Briefly, ATOCA models each pixel on the detector as a combination of contributions from each order—thereby allowing for order decontamination. That is to say, ATOCA produces decontaminated models of the first and second orders, on which a box extraction routine can be safely performed. Critically, ATOCA relies on an estimate of the underlying spatial profile of each order to accurately perform the decontamination. For clarity, we will refer to the trace profile assumption used by ATOCA as the "Reference Trace," the decontaminated trace produced by ATOCA as the "Modeled Trace," and the observed data itself as the "Observed Trace."

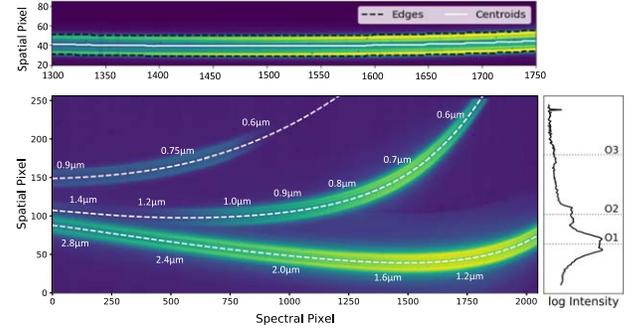

**Figure 1.** Bottom: (Left) Simulated SOSS observation (see Section 3) displaying the first three diffraction orders. The trace centroids for each order, as determined via the "edgetrigger" algorithm, are denoted with the white dashed lines, and the approximate position of several wavelengths are marked. (Right) A vertical cut along column 750, showing the structure of the spatial profiles for each order. The centroids of each of the three orders are marked with a horizontal dotted line. The spike near the top of the column is due to a bad pixel. Top: Zoom in on the 1.2–1.8 $\mu$m (columns 1750–1300) range of the first order to show the workings of the edgetrigger algorithm (see Section 2.1). The detected edges of the trace are marked in black, and the resulting centroids in white.

Here we present the A Producer of ProfiLEs for SOSS (APPLESOSS) method to estimate the underlying spatial profiles of all three diffraction orders for a given SOSS observation. This work is laid out as follows: the APPLESOSS method is first presented in Section 2. Section 3 presents the simulated observations that we use for validation of APPLESOSS, the results of which are shown in Section 4. Lastly, we assess the effects of different extraction methods on science observables (i.e., atmospheric metallicity, C/O ratio) by performing full atmosphere retrievals on our simulated data in Section 5. We provide a brief discussion and conclude in Section 6.

## 2. The APPLESOSS Method

In order to function optimally, ATOCA requires as input several reference files, including the 2D detector wavelength solution, the instrument throughput (also referred to as the photon conversion efficiency), and the spatial profiles for each order. The wavelength solution and throughput will be measured and made available during commissioning. APPLE-SOSS is a publicly available tool[12] that was developed to provide the final critical piece to the ATOCA algorithm—the spatial profiles for each order. It is designed to be data driven, and only employ modeling when absolutely necessary. Most of the SOSS trace, both of order 1 and order 2 (there are currently no plans to extract order 3), is uncontaminated, and we can therefore effectively "read off" the shape of the spatial profile. Only a comparatively small region is contaminated, thereby requiring a more delicate treatment—and even within this region we endeavor to reuse as much of the original data as possible.

---

[11] The wavelength range NIRSpec Prism mode also extends down to 0.6 $\mu$m, but at a much lower resolution of $R \sim 30$, compared to $R \sim 700$ for SOSS.

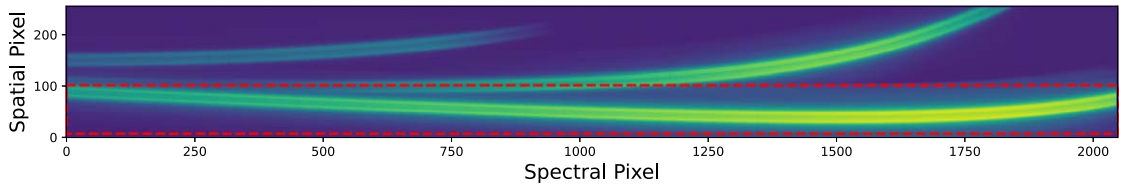

**Figure 2.** Same as the middle panel of Figure 1, showing the *x* and *y* axes with equal scaling to better visualize the detector. The subarray for bright star observations, SUBSTRIP96, is denoted with the dashed red box.

APPLESOSS can be broadly separated into two parts: centroiding, and the reconstruction of orders 1 and 2 (and/or 3), both of which will be described below. The only input that is required is a high signal-to-noise SOSS observation. Currently APPLESOSS is able to reconstruct models of all three orders for the FULL and SUBSTRIP256 subarrays, but only order 1 for SUBSTRIP96.

### 2.1. Centroiding via the EdgeTrigger Method

The first step to obtaining the shape of the spatial profile is determining the location of each order, via its centroid, as accurately as possible. One possible method for locating the centroid of the SOSS point-spread function (PSF) for a given order along the spatial axis would simply be via a "center-of-mass" analysis. However, in the detector region where orders 1 and 2 overlap, the resulting centroid is shifted toward the contaminating signal and that technique fails, even when masking the contaminating signal. To resolve this problem, we therefore chose to define the trace centroid as the midpoint between its two edges, and created an algorithm, which we term the "edgetrigger" method, to perform this calculation.

Briefly, the edgetrigger algorithm locates the edges of the trace of the target order by finding the extrema (maximum on one side, minimum on the other) of its derivative along the spatial axis. The centroid is then simply the midpoint between these two edges. For orders whose trace is partially blended, it is sufficient to measure a single edge position (the bottom edge of order 1 or top edge of order 2). The midpoint can then be retrieved based on our empirical knowledge of the trace width as a function of wavelength which was obtained from the uncontaminated regions.

We note that the profile signal smoothly rises along its edges over several ($\gtrsim$5) pixels while the core of the profile harbors variations at smaller spatial scales. By adopting a smoothed derivative (over $\pm$5 pixels), we ensure that this measurement is robust against high frequency variations in the profile.

To further improve robustness, masks specific to each spectral order are applied to ensure that only two edges (or one edge) along the profile are present (e.g., orders 2 and 3 are masked out when locating the edges of order 1; orders 1 and 2 are masked out when finding the edges of order 3, etc.). For order 2, the edgetrigger method combines a two-edge approach blue-wards of column 500 (free of order overlap) with a one-edge approach redwards (with order overlap). In all cases, after centroids are determined at each detector column, a polynomial is fit to even out any local centroid deviation. On sky, we expect additional masking tailored to each target to be necessary to deal with spectral traces originating from nearby field stars due to the slitless nature of the SOSS mode.

The resulting centroids for all three orders are shown in the white dashed lines in the lower left panel of Figure 1, and can be seen to accurately trace the position of the trace. We also note here that before running the edgetrigger algorithm, pixels that were previously flagged as bad-pixels are interpolated using the median of a box of neighboring pixels.

### 2.2. Reconstruction of the Orders

The first order is the brightest of the three orders on the detector, with its signal nearly overwhelming that of the other two orders. After determining the centroid positions, APPLE-SOSS therefore starts by producing a model of the first order spatial profile. As mentioned above, the goal of APPLESOSS is to reuse as much of the original data as possible. This goal is made possible by the fact that the cores of orders 1 and 2 (as well as 3) do not completely overlap across the whole detector —the wings of all three orders certainly suffer from contamination, however the cores remain mostly distinct. This means that for most detector columns, the core of each order

---

[12] https://github.com/radicamc/applesoss

can be effectively "read-off" of a data frame, and only the wings need to be reconstructed.

The wing reconstruction is facilitated by the `WebbPSF` python package (Perrin et al. 2014), which allows for the simulation of realistic PSFs for each of the four JWST instruments. Currently, `WebbPSF` uses a best estimate realization of the telescope wave front error (WFE) map, as well as full optical models for each instrument to generate appropriate PSFs. For SOSS, this notably includes the GR700XD grism as well as the de-focusing cylindrical lens. However, an on-sky WFE map will soon be implemented using the results of commissioning observations to much more accurately reflect the on-sky PSFs of each instrument.

A series of spatial profiles are thus simulated across the full SOSS wavelength range. As this process can be quite time consuming, we elect to construct a discrete grid of PSFs, at a default wavelength spacing of 0.1 $\mu m$. This grid spacing can be modified by the user, however we find 0.1 $\mu m$ to be the optimal tradeoff between speed and accuracy—the grid can be generated in a matter of minutes, and the use of finer spacings results in accuracy improvements of only a fraction of a percent (as quantified by the decontamination rms; see Figure 6).

For a given detector column, we obtain the wavelength of the first order at that column via the wavelength solution, as well as the centroid position of the trace via the edgetrigger method. We then shift (via spline interpolation) the two nearest nodes of the `WebbPSF` grid to the correct centroid position, and linearly interpolate to construct an estimate of the profile at the wavelength in question. The core of this new profile is then masked, and a ninth order polynomial is fit to either wing.

To perform the final reconstruction, the two simulated wings are scaled up to the flux level of the first order profile. The core of the first order, taken from the original data, is then stitched to the two wings via a linear interpolation over the pixels surrounding the break point to ensure smoothness. A visualization of this process is shown in Figure 3.

The same process is then followed to reconstruct the second and third orders. For the second order, the reconstructed first order profile is initially subtracted off of the detector, leaving the second order as the dominant contribution. As described above, the core of the profile is reused, and stitched to simulated wings. Complications arise however for the region of order 2 corresponding to wavelengths $\gtrsim 1.1$ $\mu m$ (pixel index $\lesssim 700$), where the core of the second order begins to be buried within order 1. In this case it is impossible to reuse even the core of the second order. However, this is also the region where the wavelength domain is common to the first and second order. The shape of the PSF is completely determined by the telescope optical system, and is fixed for a given wavelength—that is the shape of the spatial profile at a certain wavelength does not depend on the order in which it is located. Since we have already reconstructed profiles for the $\sim 1.1$–$1.4$ $\mu m$ range in the first order, we can simply reuse these to reconstruct the second order in this region.

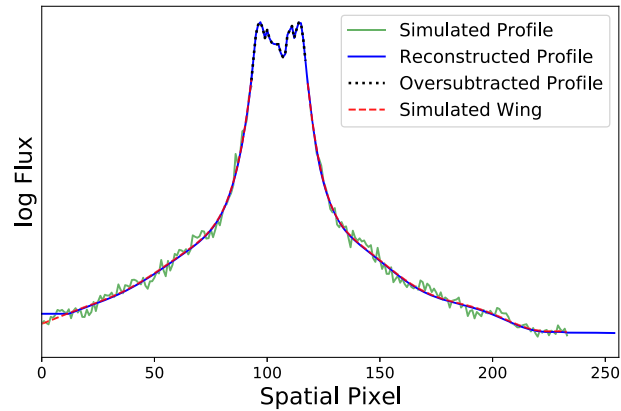

**Figure 3.** Visualization of the spatial profile reconstruction—shown here for the second order, but the same process applies for all orders. The core of the original second order profile after the subtraction of the modeled first order, for column 1000 (wavelength $\sim 0.9$ $\mu m$) is shown in the black dotted line. As the wings are much lower level than those of the first order, they have almost been completely subtracted off. The profile simulated with the `WebbPSF` package is shown in green. The wings extracted from the simulated profile are shown in the red dashed line, and the final reconstructed profile is in blue.

To reconstruct the second order profiles for pixel indices $\lesssim 700$, we thus locate the column in order 1 with the same wavelength as the order 2 column in question. As the second order has a higher spectral sampling than order 1, profiles generally need to be interpolated from the two closest wavelengths in order 1. Here, we once again use a linear interpolation as the profile (even the core) will not change greatly over the span of a thousandth of a micron. The first order profile is then shifted (via spline interpolation) to the centroid position of the corresponding wavelength in the second order. The above process is then repeated for the third order after subtracting both the first and second order reconstructions.

After the reconstruction of all three orders, APPLESOSS performs some minor post-processing to adapt the modeled spatial profiles to the form expected by ATOCA. Namely, a user specified amount of padding can be added to the spatial and spectral axes for each order, and the profiles can be oversampled if requested. Lastly, the profiles are normalized such that each column sums to one—as is required by ATOCA. The profiles are then written to a file in the correct format to be ingested directly into ATOCA as the `specprofile` reference file. An example of the final spatial profiles for each order are shown in Figure 4, and a comparison between the APPLE-SOSS reconstructions and the true underlying spatial profiles is shown in Figure 5.

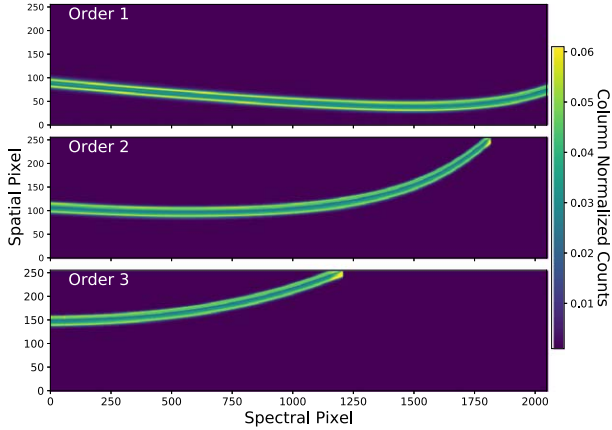

**Figure 4.** Example spatial profiles constructed with APPLESOSS for the first, second, and third SOSS diffraction orders. The profiles are highly data-driven, requiring only a high signal-to-noise SOSS observation.

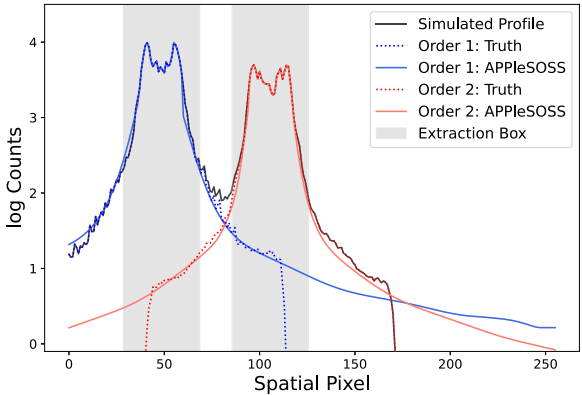

**Figure 5.** Visualization of the performance of APPLESOSS. In black is the spatial profile of column 1000 ($\sim 2\,\mu$m in order 1 and $\sim 0.9\,\mu$m in order 2). The dotted blue and red profiles are the "true" profiles for the first and second order respectively which were seeded in the simulation. The solid blue and red lines show the APPLESOSS reconstructions (after scaling back up to the native counts level). The faded gray boxes indicate the areas considered in the box aperture extraction after decontamination. Note that the spatial profiles used as input to the simulation were artificially truncated after a width of $\sim 100$ pixels.

## 3. Simulation and Data Reduction

We now proceed to validate the performance of APPLE-SOSS when used in conjunction with the ATOCA algorithm. For the validation we use simulated TSOs produced via the Instrument Development Team SOSS simulator (IDTSOSS), the details of which will be presented in L. Albert et al. (2022, in preparation). However, we outline the major steps here for completeness.

An image is created at every time step (every integration) of a time-series simulation and formatted in the expected JWST data format. The IDTSOSS simulations use a grid of

PHOENIX stellar spectra generated by Peter Hauschildt at high resolving power ($R \geqslant 250{,}000$) on a 30-point grid of specific intensities covering the whole wavelength range of JWST (0.5–30 $\mu$m). The input stellar spectrum is combined with the wavelength-dependent planet radius based on the *transitfit5* transit model (Rowe [2016]) calculated with user-defined input orbital parameters. This is performed at the highest possible resolving power ($R \geqslant 75{,}000$). The combined spectrum is then multiplied by the wavelength-dependent instrument throughput (including JWST optics and detector quantum efficiency) anchored on the $J$-band magnitude. The resulting photon flux is seeded as a thin (1 pixel wide) trace on a $4\times$ oversampled detector grid at the detector positions calibrated during cryogenic vacuum testing of the instrument. Finally, this thin trace is convolved with a range of 100 monochromatic PSFs generated using `WebbPSF`, spanning the whole wavelength range of interest. Each spectral order is treated independently, then combined to create the final detector image.

Once a full noiseless simulation is generated, multiple realistic detector effects and sources of noise (e.g., superbias, dark current, flat field, $1/f$ noise, nonlinearity, etc.) can be added to the simulation to more accurately represent a true SOSS observation. However, since our goal is to assess the quality of the extraction, we choose to keep our simulations noiseless. We can therefore be sure that any biases, or differences in our extracted spectrum compared to the input atmosphere model come from the extraction itself, and not improperly corrected detector effects. We note however, that the results presented below hold regardless of whether we use noisy, or noiseless simulations.

For this study, we select WASP-52 b, a 1.27 $R_{\rm Jup}$, 0.46 $M_{\rm Jup}$ inflated hot-Jupiter (Hébrard et al. [2013]) orbiting a 0.85 $M_\odot$ K dwarf as our test case. WASP-52 b is slated to be observed during Cycle 1 of JWST observations as part of the NEAT GTO program.[13] For the input planet atmosphere, using the SCARLET atmosphere framework (Benneke & Seager [2012], [2013]; Benneke [2015]), we simulated the cloud-free transmission spectrum of WASP-52 b assuming an atmosphere in chemical equilibrium with a metallicity of $100\times$ solar and a solar C/O ratio (0.54). A deep stack of the resulting simulated WASP-52 b SOSS TSO (but with all noise sources added to improve visualization) is shown in Figure 1.

We processed the simulated TSO through only the `RampFitStep` step of the `calwebb_detector1` stage of the JWST pipeline,[14] which performs the "up-the-ramp" fitting for each integration. The other steps deal with the correction of detector effects, which are not included in our simulation.

---

[13] GTO Program 1201.
[14] https://jwst-pipeline.readthedocs.io/en/latest/jwst/introduction.html

## 4. Spectral Extraction with ATOCA

The ATOCA algorithm is implemented in the `Extract1dStep` step of the JWST pipeline `calwebb_spec2` stage. However, it is currently not the default option, and must be toggled on via specifying `soss_atoca = True` in the `Extract1dStep` call.

We perform four different extractions; the first three of which utilize ATOCA and experiment with the effects of using different Reference Traces. The fourth is a box extraction ignoring the effects of contamination. Moving forward, we will refer to the four cases as follows:

1. (ATOCA: Ideal): There is a perfect match between the Reference Trace and the Observed Trace.
2. (ATOCA: Incorrect WFE): The Reference Trace has completely different WFE substructure than the Observed Trace.
3. (ATOCA APPLESOSS): The Reference Trace is generated with APPLESOSS.
4. (Contaminated Box): Contamination is ignored, and a naïve box extraction is performed.

Case #1 will evidently not be possible with real observations, as we will never know *exactly* what the underlying spatial profile is. However, since we are working with simulated observations, we can use the same WebbPSF spatial profiles that were seeded in the simulation as the Reference Trace in ATOCA. This therefore represents a "best possible" outcome for the extraction.

Conversely, Case #2 represents a "worst possible" outcome with ATOCA. Here, we simulate a Reference Trace with WebbPSF but using a completely different WFE realization than was used for the simulation. The WFE is a property of the optical system of the instrument, and though it may evolve over the lifetime of the JWST due to minute changes in the mirrors or other optical elements, it is highly unlikely to drastically change over short time frames. Small changes in the WFE can translate into significant alterations to the spatial PSF (e.g., changing the amplitude or position of the most prominent peaks, widening or narrowing the PSF, etc.). Therefore, for this test, the Reference and Observed Traces have very different fine structure, although the large-scale features remain mostly consistent (e.g., the location of the peaks).

Case #3 allows us to validate APPLESOSS. As described in Section 2, we generate traces for each order using the simulated data itself.

Case #4 is the simplest of all—ignore the contamination and just do a box aperture extraction. As mentioned above, DB22 predicted that the effects of contamination should be small for relative measurements, and this test offers an opportunity to validate that prediction.

To assess the quality of the ATOCA decontamination for Cases 1–3, we plot the residual detector image of the form:

$$(\mathrm{Data} - \mathrm{Model}_{O1} - \mathrm{Model}_{O2})/\mathrm{Error}, \qquad (1)$$

for each case. Here, $\mathrm{Model}_{O1}$, and $\mathrm{Model}_{O2}$ are the Modeled Traces produced by ATOCA for orders 1 and 2 respectively. The Error values are the flux error on individual pixels. Since our simulations are noiseless, the flux error values are all zero. However, ATOCA requires errors to be non-zero to function. We therefore inject the expected level of photon noise into the simulation and assign each pixel the corresponding photon noise-limited errors. This results in a S/N at ~1.6 $\mu$m (near the peak of the order 1 throughput) in the first order of ~250. The resulting decontamination residuals are shown in Figure 6.

For the ideal case, the residuals are nearly all within 1$\sigma$, showing that the Modeled and Observed Trace agree to within the error margin of the data—exactly what we would expect given that the Reference Trace was identical to the Observed Trace. Indeed, the rms of the residuals is 0.474. For the APPLESOSS case, the residuals for order 1 are also within 1$\sigma$, showing that we can very accurately extract the spatial profile of the first order. There are slightly more issues with the second order, especially at the reddest wavelengths where we are not able to reconstruct the second order profile directly. However, the rms is still a favorable value of 0.766—indicating that the Modeled Trace agrees in general with the Reference Trace to within the errors. Lastly, for the incorrect WFE case, the residual detector map can be clearly seen to be much worse than for the other two cases. Large (>5$\sigma$) deviations are seen throughout, and the rms of the residuals is 7.318.

No decontamination plot is produced for the Contaminated Box extraction case as no decontamination is performed. Instead, the flux is summed in a box of 40 pixels centered on the edgetrigger centroid for each column; both for order 1 and order 2.

To further contrast all four cases, we show the extracted 1D stellar spectra of an out-of-transit frame (that is where no contribution from the planet is present) in Figure 7. The spectra themselves are mostly indistinguishable for the four cases (except for long wavelengths of order 2, where the Contaminated Box case begins to diverge). By comparing the spectrum from each case to the true underlying stellar spectrum, determined by performing a box extraction on the individual seed traces produced by the IDTSOSS simulator for each order, we notice a few key discrepancies. The Contaminated Box extraction over-predicts the flux at the longest wavelengths in order 1, although this only amounts to a bias of a fraction of a percent. This over-prediction however is markedly larger for order 2. For wavelengths longer than ~1 $\mu$m in order 2, the contaminated box extraction erroneously assigns flux from order 1 to order 2, resulting in gross over-estimations of the flux of several tens of percent. This indeed matches the predictions of DB22—and confirms that proper treatment of the order contamination is essential for *measurements of absolute flux* with the SOSS mode.

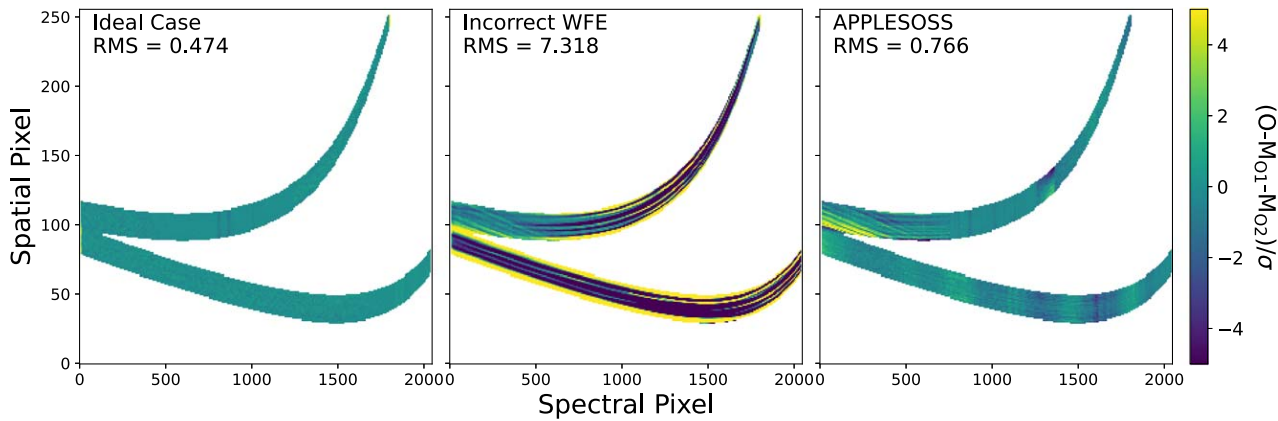

**Figure 6.** Decontamination residuals for each of the three cases using ATOCA. All pixels outside the box aperture of 40 pixels that was assumed in ATOCA are masked.

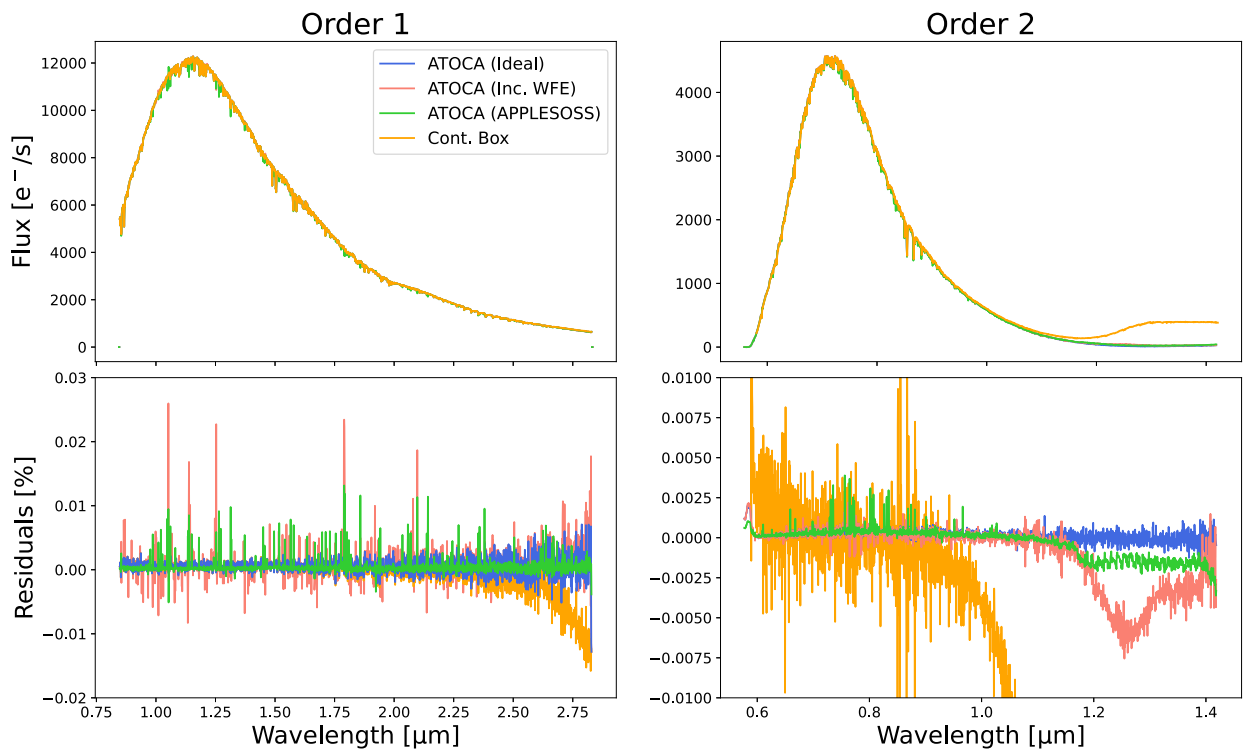

**Figure 7.** Top: Stellar spectra extracted from one out-of-transit integration of the simulated WASP-52 b TSO. The left panels show the first order, and the right panels the second order. The three ATOCA extractions are in blue (ideal case), red (incorrect WFE case), and green (APPLESOSS case), whereas the contaminated box extraction is in orange. Bottom: Percent residuals computed as the difference between the true spectrum and the extracted spectrum for each case, of the form (true–observed)/true. Several percent residuals persist for the contaminated box extraction, particularly in order 2. All three ATOCA extractions though, maintain residuals of less than a hundredth of a percent.

The three ATOCA cases all have residuals of a fraction of a percent across order 1. Some flux over-predictions are visible at the reddest end of order 2 for the Incorrect WFE and APPLESOSS cases, indicating that the decontamination is not perfect. However, these over-estimations are capped at less than a hundredth of a percent compared to the tens of percent for the Contaminated Box case. Furthermore, APPLESOSS can clearly be seen to outperform the Incorrect WFE extraction in terms of flux over-estimation in the contaminated region of order 2.

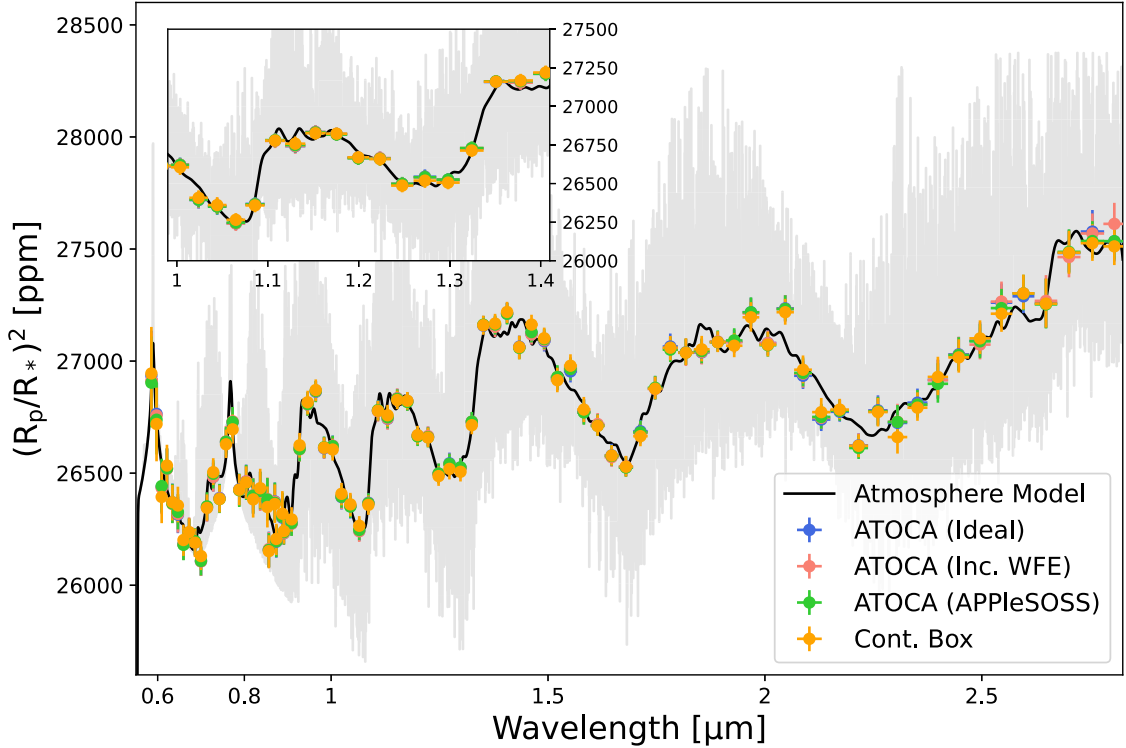

**Figure 8.** Fitted transmission spectra for each of the four test cases. The full resolution ($R \sim 10^6$) atmosphere model of WASP-52 b seeded into our simulation is shown in gray, and the model at a resolution of $R = 50$ is shown in black. The inset in the upper left corner shows a zoom-in on the 1.0–1.4 $\mu$m water bands to attempt to better visualize the differences between the four cases.

### 4.1. Light Curve Fitting

Although the decontamination results look promising, the end product that matters for exoplanet atmosphere science is the resulting transmission (or emission) spectrum, and the science parameters which can be obtained from it. Exoplanet atmosphere measurements are inherently a relative measurement, that is flux in-transit is compared to flux out-of-transit (or eclipse), and the effects of contamination are predicted to be much smaller than those presented in Figure 7. We thus proceed to fit our extracted light curves and generate a transmission spectrum for each of the four cases.

We fit the extracted light curves following standard procedures (e.g., Kreidberg et al. 2015; Benneke et al. 2019). First, we construct, and fit a transit model to the broadband white light curve. Using the flexible `juliet` package (Espinoza et al. 2019), we fit for five parameters: central transit time, $t_0$, scaled planetary radius, $R_p/R_*$, two parameters describing quadratic limb-darkening following the parameterization of Kipping (2013), as well as a scalar jitter parameter. We fix the period (1.75 days), orbital eccentricity (0), impact parameter (0.6), and stellar density (2.48 g cm$^{-3}$) to the values determined by Hébrard et al. (2013).

We then bin the extracted spectra from their nominal resolution of $R \sim 1000$ to $R \sim 50$, and via visual inspection, remove edge wavelength bins where the signal-to-noise ratio is too low to clearly display any transit-like signal. We chose $R \sim 50$ mostly for ease of visualization of the results; however these results hold irrespective of the chosen resolution. Throughout this process we maintain the first and second orders separate—ending up with 61 bins for order 1, and 30 for order 2. We employ `juliet` once again to fit each spectroscopic light curve, additionally fixing $t_0$ to its white light value. We therefore fit four parameters for each spectroscopic light curve to obtain a final transmission spectrum. The transmission spectra for each case are plotted in Figure 8.

The differences between the transmission spectra for the four cases are extremely small—indeed the points are effectively all overlapping in Figure 8. Even in the blown-up inset of the 1.4 $\mu$m water band, the differences are still nearly non-existent. In each case, the fitted transmission spectrum nearly perfectly traces the input atmosphere model. Indeed, this level of agreement, even between the ATOCA (ideal) and Contaminated Box extraction scenarios was predicted by DB22. The second order contamination acts as a dilution factor on the first order transit depth (and vice-versa) at a given detector column. However, this differs from standard dilution problems (e.g., a background star falling within the aperture during a transit observation) in that both the target order and the contaminating order "see" the transit. Therefore, the resulting dilution is only by a factor of the *difference in transit depth* between order 1 and 2 (see Equation (5) in DB22). To this end, DB22 estimate that for order 1, the effect of the second order contamination will be approximately 1% of the size of the spectral feature expected at a given wavelength. For example, since our WASP-52 b spectrum shows water band features at an amplitude of ∼500 ppm at ∼2 μm, an estimate of the second order contamination would be ∼5 ppm—in this case, less than the predicted photon noise. It is therefore not surprising that all four of our tests give nearly identical transmission spectra.

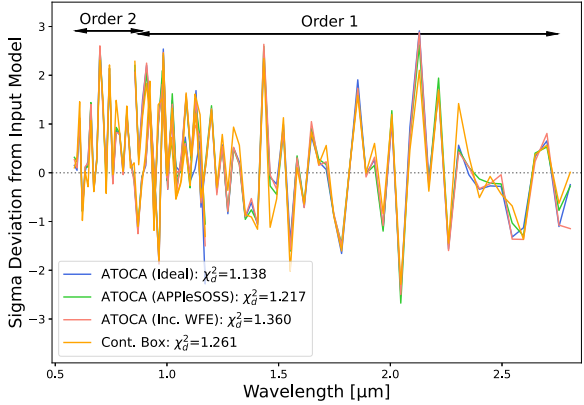

**Figure 9.** Point-to-point differences between the extracted transmission spectrum for each of the four cases and the input atmosphere model, divided by the error on the transmission point. These differences are, on average, within ∼1σ, confirming the accuracy of the transmission spectrum in each case. The $\chi_d^2$ goodness-of-fit value ($\chi^2$ divided by the number of data points) for each case is listed in the legend.

Figure 9 further quantifies the differences between the transmission spectrum in each case and the atmosphere model used as input to the IDTSOSS simulation. In general, the differences are all within 1σ (where σ is the 16th–84th percentile bounds of $(R_p/R_*)^2$ determined by `juliet`)—highlighting the fact that each extraction agrees with the input model to within the photon noise-limited errors. On-sky of course, it still remains to be seen whether SOSS can indeed achieve photon-noise limited precision. Yet, we still obtain agreement to within 1σ on simulations including the full complement of detector effects and noise sources.

Moreover, the $\chi_d^2$ ($\chi^2$ divided by the number of data points) goodness-of-fit value for each case is extremely favorable, with the best case being the ATOCA (ideal) spectrum. The ATOCA (APPLESOSS) spectrum is next, followed by the Contaminated Box spectrum, and then the ATOCA (Incorrect WFE) spectrum. However, all the $\chi_d^2$ values are between 1.1 and 1.4, indicating that on average, the model is consistent with each data point for each case at the ∼1.2σ level. There is thus, little evidence to choose between any of the four cases. There are also no systematic biases present in the point-to-point differences, rather random distributions about zero.

We therefore demonstrate that both the stellar and transmission spectra obtained using the APPLESOSS spatial profiles retain a high degree of fidelity to the input models.

## 5. Atmosphere Retrieval

Although unlikely, it is possible that small differences in transmission spectra can result in differences of several σ in retrieved atmosphere parameters—especially in high precision spectra (Barstow et al. 2020). As a final test, we therefore perform a suite of retrieval simulations utilizing the open source code CHIMERA (Line et al. 2013) for each of our four cases. We focus on the ability to correctly constrain the atmosphere C/O and metallicity as these are key parameters determining the atmosphere's composition, and allow insights into important questions such as: what are the abundances of different chemical species and where in the protoplanetary disk did the planet form?

For each transmission spectrum discussed in Section 4 we perform a chemical equilibrium spectral retrieval, where the molecular and atomic vertical abundances are assumed to be in thermochemical equilibrium. These abundances are computed using the NASA CEA (Chemical Equilibrium with Applications) model (Gordon & McBride 1994) for a given C/O and metallicity. To accommodate the overlapping wavelength regions of order 1 and order 2 we perform a joint retrieval; that is to say, the likelihood used in the retrieval is the sum of the individual likelihoods of each order, log $L_i$, such that $\log L_{\text{total}} = \sum_i \log L_i$, where

$$\log L_i = -\frac{1}{2} \sum_j \frac{(y_{i,j} - y_{\text{mod},i,j})^2}{\sigma_{i,j}^2}. \qquad (2)$$

In Equation (2), $y_{i,j}$ are the observed data points, $y_{\text{mod},i,j}$ are the modeled data points obtained from the retrieval analysis, and $\sigma_{i,j}$ are the errors on the observed data points.

The C/O and metallicity posterior distributions for each case are shown in Figure 10. For each, the retrieved C/O ratio is within 1σ of the input value, and the retrieved metallicity is within ∼1.2σ. The comparatively more uncertain determination of metallicity can be attributed in large part to the wavelength regime of the SOSS mode. SOSS is predominantly sensitive to the NIR water bands in the 0.9–2.5 μm range, and it is difficult to estimate the bulk metallicity of an atmosphere from measurements of water alone. The metallicity constraint is aided by the presence of alkalis at short wavelengths, as well as $CO_2$ and CO features >2.5 μm. However, the most precise metallicity measurements will likely come from combining SOSS observations with other instruments providing longer wavelength coverage (e.g., the NIRSpec G395 grism).

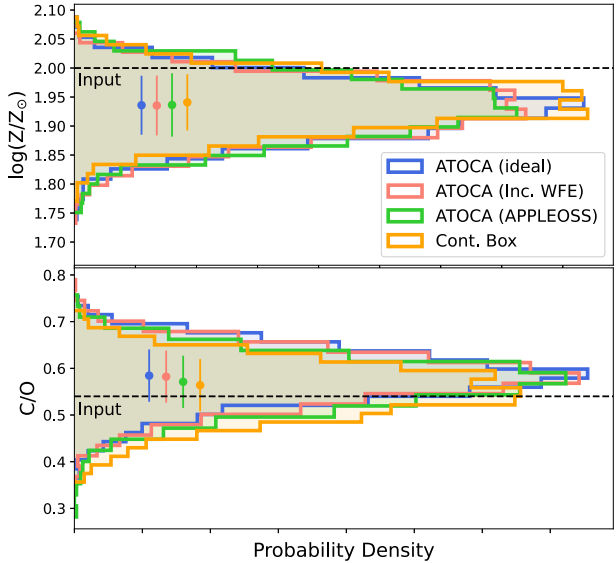

**Figure 10.** Posterior distributions for metallicity (top) and C/O (bottom) obtained through our CHIMERA chemical equilibrium retrievals for each of the four extractions. The posterior median, as well as 1σ error envelopes for each case are denoted by the colored error bars. Each extraction yields the input C/O within 1σ and the input metallicity within ∼1.2σ.

## 6. Discussion and Conclusions

In this work we have presented APPLESOSS, an algorithm to produce an estimate of the underlying spatial profiles for a NIRISS/SOSS observation that is required by the ATOCA extraction algorithm. We then proceeded to validate APPLE­SOSS and compare it to three other extractions using simulated SOSS transmission observations of WASP-52 b. We first compare the extracted stellar spectra for each case, and find that the APPLESOSS case yields the closest match to true stellar spectrum (other than the, unrealizable in real life, ideal case). We then compare the transmission spectra of each of four cases to the input atmosphere model in both a $\chi^2$ sense, as well as with a full atmosphere retrieval.

The results of the stellar spectrum comparison highlight the importance of properly treating the order contamination for *measurements of absolute flux*. On the other hand, for *relative measurements* of exoplanet atmospheres, there is much less differentiation. Each of our four extractions (ATOCA (Ideal), ATOCA (Incorrect WFE), ATOCA (APPLESOSS), and the Contaminated Box extraction) all result in a transmission spectrum that is nearly identical to the atmosphere model that was used as input in our noiseless simulations. Although the APPLESOSS spectrum had the second lowest decontamination residual rms of the ATOCA extractions, as well as the second lowest $\chi_d^2$ value of the four tests cases (second to the ATOCA (Ideal) extraction, as expected, in both cases) the differences between all four spectra are small (∼1σ), and no systematic biases are found in any case.

Furthermore, the retrieved C/O in each case are also within 1σ of the true values used in our simulation, and the metallicity with ∼1.2σ. The retrieved error envelopes are also similar in each of the four cases, with no one case yielding results of a higher or lower precision than the others.

In all, we can draw three major conclusions from our tests:

1. As predicted by DB22, although the effects of the order contamination can be significant for measurements of absolute flux, for relative measurements (e.g., exoplanet transmission observations) the effect is quite small. Indeed, Case #4, the Contaminated Box extraction, which ignored the effects of the contamination yields results nearly identical to the other three ATOCA extractions.

2. The ATOCA algorithm is quite robust against errors in the input Reference Trace. Even though a bad Reference Trace (e.g., Case #3, the Incorrect WFE) yields a concerning decontamination plot (see Figure 6), the transmission spectrum resulting from this extraction, as well as the retrieved atmosphere parameters are unaffected.

3. Regardless of how robust ATOCA is to differences in the Reference Trace, *a* Reference Trace is nonetheless required. We demonstrate that APPLESOSS traces retain a high degree of fidelity to an observed SOSS observation, and are therefore the best candidate for this job.

Indeed, further to this final point, APPLESOSS has been applied to a NIRISS/SOSS TSO as part of commissioning, resulting in the `specprofile` reference file which is included in the JWST pipeline. However, in case there are changes in the PSF from observation to observation, it should be preferable to generate a trace profile for each individual TSO. To this end, we make APPLESOSS publicly available at http://github.com/radicamc/applesoss to allow the community to take maximal advantage of the ATOCA algorithm.

A potential parallel path to differentiating the spatial profiles of the first and second orders in the contaminated region is provided via the availability of the F277W filter. The F277W filter, when used in combination with the G700XD grism instead of the standard CLEAR filter, limits the SOSS wavelength range to 2.4–2.8 μm—effectively isolating the contribution of the first order in the contaminated region. For programs which have included such exposures, the F277W filter may provide an alternate avenue construct Reference Traces, and such an option will likely be implemented in APPLESOSS in the near future.

One further aspect of the SOSS mode that has not been considered so far is the effect of contamination from background sources. Since SOSS is a slitless observing mode, contributions from field stars can also contaminate the trace of the target star. By suitably tuning the telescope orientation, such contamination can be mostly mitigated. However, for targets where background contamination is unavoidable, APPLESOSS could provide a way forward to model and remove these contaminating background traces, and further tests and alterations to the algorithm will be implemented once on-sky data is widely available.

M.R. would like to acknowledge funding from the Natural Sciences and Research Council of Canada (NSERC), as well as from the Fonds de Recherche du Québec—Nature et Technologies (FRQNT) and the Institut de Recherche sur les Exoplanètes (iREx) for support toward his doctoral studies, as well as GoGo squeeZ for making quality applesauce. D.J. is supported by NRC Canada and by an NSERC Discovery Grant. C.P. acknowledges financial support from FRQNT, NSERC, as well as the Technologies for Exo-Planetary Science (TEPS) Trainee Program. This project was undertaken with the financial support of the Canadian Space Agency.

*Software:* astropy; Astropy Collaboration et al. (2013, 2018), ipython; Pérez & Granger (2007), juliet; Espinoza et al. (2019), matplotlib; Hunter (2007), numpy; Harris et al. (2020), pymultinest; Buchner (2016), scipy; Virtanen et al. (2020), WebbPSF; Perrin et al. (2014).

## ORCID iDs

Michael Radica  https://orcid.org/0000-0002-3328-1203
Loïc Albert  https://orcid.org/0000-0003-0475-9375
Jake Taylor  https://orcid.org/0000-0003-4844-9838
David Lafrenière  https://orcid.org/0000-0002-6780-4252
Louis-Philippe Coulombe  https://orcid.org/0000-0002-2195-735X
Antoine Darveau-Bernier  https://orcid.org/0000-0002-7786-0661
René Doyon  https://orcid.org/0000-0001-5485-4675
Neil Cook  https://orcid.org/0000-0003-4166-4121
Nicolas Cowan  https://orcid.org/0000-0001-6129-5699
Néstor Espinoza  https://orcid.org/0000-0001-9513-1449
Doug Johnstone  https://orcid.org/0000-0002-6773-459X

Lisa Kaltenegger  https://orcid.org/0000-0002-0436-1802
Caroline Piaulet  https://orcid.org/0000-0002-2875-917X
Arpita Roy  https://orcid.org/0000-0001-8127-5775
Geert Jan Talens  https://orcid.org/0000-0003-4787-2335